\documentclass[a4paper,11pt]{article}
\usepackage[margin=2.5cm]{geometry}
\usepackage{setspace,graphicx,url,hyperref,enumitem}
\usepackage{amsmath,amssymb,amsfonts,amsthm}
\usepackage{color,colortbl}
\definecolor{Gray}{gray}{0.9}
\newcommand{\N}{\ensuremath{\mathbb{N}}}
\newcommand{\R}{\ensuremath{\mathbb{R}}}

\begin{document}
\onehalfspacing
\title{Optimizing embedding-related quantum annealing parameters for reducing hardware bias}
\author{Aaron Barbosa, Elijah Pelofske, Georg Hahn, and Hristo N.\ Djidjev}
\date{Los Alamos National Laboratory}
\maketitle

\begin{abstract}
    Quantum annealers have been designed to propose near-optimal solutions to NP-hard optimization problems. However, the accuracy of current annealers such as the ones of D-Wave Systems, Inc., is limited by environmental noise and hardware biases. One way to deal with these imperfections and to improve the quality of the annealing results is to apply a variety of pre-processing techniques such as spin reversal (SR), anneal offsets (AO), or chain weights (CW). Maximizing the effectiveness of these techniques involves performing optimizations over a large number of parameters, which would be too costly if needed to be done for each new problem instance. In this work, we show that the aforementioned parameter optimization can be done for an entire class of problems, given each instance uses a previously chosen fixed embedding. Specifically, in the training phase, we fix an embedding $E$ of a complete graph onto the hardware of the annealer, and then run an optimization algorithm to tune the following set of parameter values: the set of bits to be flipped for SR, the specific qubit offsets for AO, and the distribution of chain weights, optimized over a set of training graphs randomly chosen from that class, where the graphs are embedded onto the hardware using $E$. In the testing phase, we estimate how well the parameters computed during the training phase work on a random selection of other graphs from that class. We investigate graph instances of varying densities for the Maximum Clique, Maximum Cut, and Graph Partitioning problems. Our results indicate that, compared to their default behavior, substantial improvements of the annealing results can be achieved by using the optimized parameters for SR, AO, and CW.
\end{abstract}

\section{Introduction}
\label{sec:intro}
Quantum annealers such as the ones designed by D-Wave Systems, Inc.~\cite{DWaveOverview} are able to find approximate solutions to NP-hard problems of very high quality by resorting to a technique called quantum annealing. To be precise, the D-Wave annealer is designed to solve optimization problems requiring the minimization of a function of the form
\begin{align}
    H(x_1,\ldots,x_n) = \sum_{i=1}^N h_i x_i + \sum_{i<j} J_{ij} x_i x_j,
    \label{eq:hamiltonian}
\end{align}
where the linear weights $h_i \in \R$ and the quadratic couplers $J_{ij} \in \R$ are specified by the user and define the problem, and $x_i$ are unknown binary variables, where $i,j \in \{1,\ldots,n\}$. To minimize eq.~\eqref{eq:hamiltonian}, the D-Wave annealer maps the connectivity of the logical qubits in eq.~\eqref{eq:hamiltonian}, i.e., the graph defined by the set of edges $(i,j)$ for which $J_{ij}\neq 0$, to the qubits and links between them on its hardware chip, called a \textit{Chimera} graph (see \cite{Chapuis2019} for a graphical representation of the Chimera graph). The process of submitting a problem to a D-Wave machine is as follows:

\begin{enumerate}
    \item The problem of interest must be represented as the minimization of a function of the form of eq.~\eqref{eq:hamiltonian}. The function of eq.~\eqref{eq:hamiltonian} is called a \textit{QUBO} (quadratic unconstrained binary optimization) problem if $x_i \in \{0,1\}$ for $i \in \{1,\ldots,n\}$, and an \textit{Ising} problem if $x_i \in \{-1,+1\}$ for $i \in \{1,\ldots,n\}$. Both QUBO and Ising formulations are equivalent \cite{Chapuis2019}. We can represent the function of eq.~\eqref{eq:hamiltonian} as a graph $P$ itself having $n$ vertices, one for each variable $x_i$, $i \in \{1,\ldots,n\}$. In this representation, each vertex $i$ is assigned a vertex weight $h_{i}$, and each edge between vertices $i$ and $j$ is assigned the edge weight $J_{ij}$.
    \item Next, the problem graph $P$ is mapped onto the hardware of the D-Wave 2000Q annealer. Since it is usually not the case that the structure of the graph $P$ perfectly matches the structure of the Chimera graph of the D-Wave 2000Q, a \textit{minor embedding} of $P$ onto the Chimera graph has to be computed. In such an embedding, some logical qubits in eq.~\eqref{eq:hamiltonian} become a \textit{chain}, which is a set of hardware qubits on the chip linked together in a way that prompts them to take the same value at the end of the anneal. Defining the chains requires the specification of a parameter determining the strength of the coupling between the qubits in a chain (the \textit{chain strength} or \textit{chain weight}). The minor-embedded problem $P$ onto the Chimera graph corresponds to a new graph $P'$, which is a subgraph of the Chimera graph.
    \item At the start of the annealing process, the qubits used in the embedding of $P'$ onto the D-Wave hardware are initialized in an equal superposition \cite{DWaveOverview,DWaveManual}. During annealing, the system is slowly driven from the neutral transverse field Hamiltonian to the user-specified QUBO or Ising problem $H$ of eq.~\eqref{eq:hamiltonian} while remaining, in theory, in the ground state.
    \item Since all qubits in a chain represent one logical qubit, they act as one in theory. However, this is not guaranteed in practice, and hardware qubits in a chain might not always take the same value after annealing. In this case, we speak of a \textit{broken chain}. There is no unique way to assign a definite value to each logical qubit employed in eq.~\eqref{eq:hamiltonian} based on its broken chain, and D-Wave offers several default methods to \textit{unembed} chains, i.e., to decide on the value to be assigned to broken chains.
\end{enumerate}

In practice, several sources of error potentially decrease the quality of the solution returned by the D-Wave annealer. First, before annealing, the linear weights and quadratic couplers in eq.~\eqref{eq:hamiltonian} have to be mapped to electrical currents on the hardware chip using a linear-to-analog converter \cite{DWaveManual}. This conversion works with a finite precision of 8 bits, thus necessarily resulting in weights spanning a range larger than 8 bits to be mapped imprecisely due to rounding errors. Moreover, so-called \textit{leakage} may occur on the physical chip from the coupler $J_{ij}$ to the adjacent linear weights $h_i$ and $h_j$, where $i,j \in \{1,\ldots,n\}$. This can likewise alter the linear weights $h_i$ and $h_j$ \cite{dwave_gauge}, where the effect is reported to be more serious for chained qubits.

One simple way to mitigate such hardware biases is the so-called \textit{spin reversal} (SR) or \textit{gauge transform}. Spin reversal works on Ising problems and is based on the idea that although, theoretically, quantum annealing is invariant under a gauge transformation (i.e., the reversal of spin-up and spin-down in a quantum system), the D-Wave annealer is not a closed system and thus breaks gauge symmetry. As a consequence, two Ising problems in which certain spins have been flipped result in (slightly) different systems when mapped onto the annealer. Solving several Ising problems with a certain number of spin reversed qubits allows us to average results, and balance out errors. In practice, we select an arbitrary subset of variables $S \subseteq \{1,\ldots,n\}$ in an Ising problem, and substitute the corresponding variables as $x_i \rightarrow -x_i$ for all $i \in S$ in the Ising problem (the corresponding linear terms and quadratic couplers have to be modified as well). This is equivalent to re-interpreting an up spin as a down spin and vice versa, thus leaving the ground state of the Ising problem invariant, but having the potential to reduce analog and systematic errors on the device. The spin reversal can be applied on two different levels, either before or after embedding eq.~\eqref{eq:hamiltonian} onto the hardware (see Section~\ref{sec:methods}). In this work we explore both variants.

Second, in theory, all qubits evolve simultaneously during the anneal process, experiencing equal changes to the tunneling energy and equally contributing to the classical energy function \cite{DWaveAnnealOffsets}. In practice, however, qubits freeze out at different times during the anneal \cite{spinreversal}, which might bias the qubit states at readout after annealing. To this end, D-Wave offers to set \textit{anneal offsets} (AO) for all individual qubits with the aim to improve the solution quality. In order to synchronize the evolution of the qubits, the D-Wave 2000Q device offers the ability to delay or advance the evolution of individual qubits within predefined ranges, meaning that qubits can individually be set to start their anneal process earlier or later compared to the default schedule. We consider setting individual anneal offsets for all qubits in this work. Analogously to spin reversal, we consider applying anneal offsets in two different ways, either using separate AO for each individual qubit, or using the same AO for all qubits in each chain (see Section~\ref{sec:methods}).

Third, all couplers on the D-Wave hardware require the specification of a weight, and as such, when embedding a logical qubit as a chain on the D-Wave hardware, couplers have to also be assigned to all pairwise connections of chained qubits. The chain weight is typically set by D-Wave, in which case some overall weight is equally distributed to all couplers in a chain. However, it can also be set manually. We read out the total chain weight (CW) determined by D-Wave for each chain, and aim to re-distribute it along the chain in an optimal way.

Although SR, AO, and CW can be effective for removing hardware biases in the annealer, it is non-trivial to optimally select the actual qubits to which a spin reversal is applied, or the values of the anneal offsets or chain weights for each new problem instance being solved. This is because each technique has around 2000 degrees of freedom. Previous work has improved upon the spin reversal transform by using classical optimization in order to find an optimal set of qubits to spin reverse \cite{SpinReversalOpt2019}. Concerning the anneal offsets, D-Wave reports that longer chains are likely to freeze out sooner during the anneal process due to their lower effective tunneling energy \cite{DWaveAnnealOffsets}, and they recommend delaying the evolution of those qubits which will be subjected to strong magnetic fields relative to the other working qubits. Moreover, \cite{King2016} suggest to advance qubits in their evolution if their final state does not contribute to the energy of the classical solution.

Since tuning all qubits for an application of SR, AO, or CW individually for each new problem instance under consideration is infeasible, we propose a different approach in this work. We aim to optimize SR, AO and CW with respect to a whole class of input problems. We carry out all optimizations in the classical set-up of training and validation sets for three NP-hard problems, the Maximum Clique, Maximum Cut, and Graph Partitioning problems. Using a differential evolution optimizer, we tune the average performance of the spin reversal transform, anneal offsets, or chain weights across all of the training graphs, and evaluate our optimized sets of parameters on a set of test graphs.

The article is structured as follows. In Section~\ref{sec:methods}, we describe the background of SR, AO, and CW, as well as the optimization framework we employed. We also introduce the NP-hard problems we consider (Maximum Clique, Maximum Cut, and Graph Partitioning). Experimental results are reported in Section~\ref{sec:experiments}. The article concludes with a discussion in Section~\ref{sec:discussion}.

\section{Methods}
\label{sec:methods}
In this section, we justify our approach of using a fixed embedding for optimizing SR, AO, and CW (Section~\ref{sec:fixed_embedding}). We provide more details on the two types of spin reversal we apply (Section~\ref{sec:spin_reversal}), as well as on anneal offsets (Section~\ref{sec:anneal_offsets}) and chain weights (Section~\ref{sec:chain_weights}). Section~\ref{sec:np_problems} defines the NP-hard problems we consider. A description of the optimization we perform to tune the application of SR, AO, and CW is given in Section~\ref{sec:optimization}.

\subsection{Using a fixed embedding }
\label{sec:fixed_embedding}
Using the same (fixed) embedding is a key ingredient of our approach. If we want the same set of optimized parameters to work for multiple problems, we need at least one invariant, and it is, in this case, the hardware embedding. We use the fact that, for an NP-hard problem, the limiting factor for embedding its problem graph $P$ is the largest complete graph that can be embedded onto the quantum annealing hardware. Hence, instead of using an arbitrary embedding of a complete graph that the D-Wave's embedding method \textit{minorminer} would randomly find, we can use a fixed one, and optimize the hardware related parameters using that fixed embedding. In addition, since we will use the same embedding many times, it makes sense to choose one with as good properties as possible. For this reason, we try several complete graph embeddings and choose one that gives the best performance overall, i.e., the best QUBO/Ising value when using default D-Wave parameters, separate for each of the three considered problems.

\subsection{Spin reversal}
\label{sec:spin_reversal}
Suppose we are given an Ising problem in the form of eq.~\eqref{eq:hamiltonian} and a set $S \subseteq \{1,\ldots,n\}$ of spins to be reversed. To transform a particular $x_i$ from $-1$ to $+1$, while keeping the value of eq.~\eqref{eq:hamiltonian} unchanged, we define a new function $H'$ with $h_i' \rightarrow -h_i$ and $J_{ij}' \rightarrow -J_{ij}$, $J_{ji}' \rightarrow -J_{ji}$ for all $i \in S$, where $j \in \{1,\ldots,n\}$. Note that the ground state energies of $H$ and $H'$ are identical, and each minimum of $H'$ is a minimum of $H$ with the $i$th variable having a flipped sign. To apply spin reversal to a set $S$, we apply the above transformation to each $i\in S$.

It is not obvious how the set $S$ should be chosen to maximize the benefit of the spin reversal. As remarked in \cite{dwave_gauge}, reversing too few spins leaves the Ising model almost unchanged, whereas applying spin reversal to too many qubits likely results in many pairs of connected qubits being transformed, thus effectively leaving the corresponding quadratic couplers unchanged. In both cases, the spin reversal transform might only have little effect. Hence, the default spin reversal implemented by D-Wave flips roughly half of the qubits randomly.

When optimizing the spin reversal for a particular problem, we are thus asked to determine for each involved qubit a binary spin reversal indicator, denoting if a particular qubit is spin reversed or not.

Note that spin reversal can be applied on two levels: First, we can flip the logical qubits in the formulation of eq.~\eqref{eq:hamiltonian}. This will be referred to as \textit{spin reversal on the chain level}, since in this case qubits which are being mapped onto the hardware as chains are either all reversed or all non-reversed. Second, we can embed the original problem graph $P$ (see Section~\ref{sec:intro}) and read out the embedded Ising problem  in the graph representation $P'$. The resulting Ising problem likely consists of more qubits, due to some logical qubits being mapped to a set of physical qubits of the D-Wave hardware, and we can flip each hardware qubit individually. This will be referred to as \textit{spin reversal on the qubit level}.

\subsection{Anneal offsets}
\label{sec:anneal_offsets}
Typically, all hardware qubits being used in a problem embedded onto the D-Wave quantum chip undergo the same anneal process simultaneously. In such a case, it is normal that qubits freeze out at different times during the anneal \cite{spinreversal}, which, however, might affect negatively the dynamics of the annealing and prevent the system from reaching its ground state. To this end, in the newest generation of the annealer, the start of the anneal process of individual qubits can be moved forward or backward in time, within specified limits.

For D-Wave 2000Q, individual anneal offsets can be specified for each hardware qubit using the parameter \textit{anneal\_offsets}, where the value $0$ denotes no offset, and positive (negative) values indicate that a qubit's anneal begins ahead of (behind) the standard schedule. The range of the variable \textit{anneal\_offsets} is machine dependent, and can be queried with \textit{anneal\_offset\_ranges}. Moreover, anneal offsets are discrete, with a machine dependent step size specified in \textit{anneal\_offset\_step}. We tune the anneal offset of each involved qubit with an optimization within the range \textit{anneal\_offset\_ranges} (given as boundary condition to the optimizer) over a discrete search space, similar to spin reversal optimization.

\subsection{Bias distribution on physical qubits}
\label{sec:chain_weights}
\begin{figure}[t]
    \centering
    \includegraphics[width=0.9\textwidth]{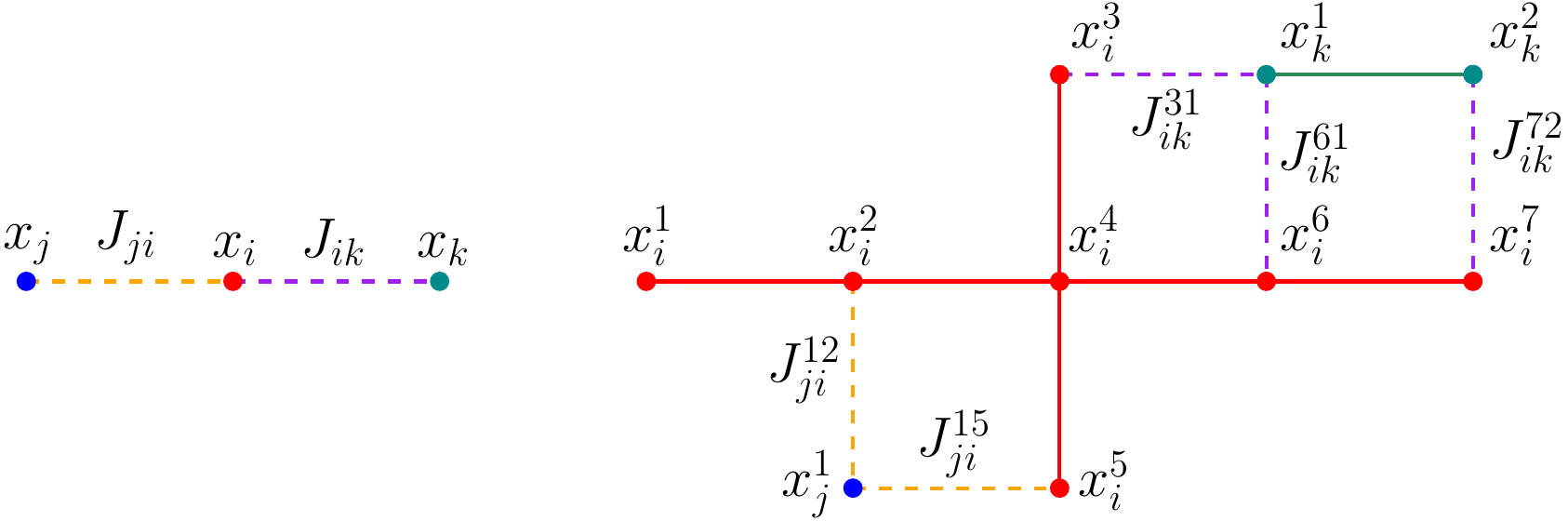}
    \caption{The logical qubit $x_i$ (left) is mapped onto a chain of seven physical qubits $x_i^1,\dots,x_i^7$ (right), and the logical couplers $J_{ji}$ and $J_{ik}$ (connecting $x_i$ to qubits $x_j$ and $x_k$) are mapped to the physical couplers $J_{ji}^{15}$, $J_{ji}^{12}$ and $J_{ik}^{31}$, $J_{ik}^{61}$, $J_{ik}^{72}$, respectively.\label{fig:chain_biases}}
\end{figure}
Typically, logical qubits have to be mapped to chains of hardware qubits when computing a minor embedding of a QUBO/Ising problem of eq.~\eqref{eq:hamiltonian} onto the D-Wave Chimera graph. In this case, a logical qubit (variable) $x_i$ is mapped onto a chain $\{x_i^1,\ldots,x_i^l\}$ having $l \in \N$ hardware qubits. The linear bias $h_i$ and the quadratic biases $J_{ij}$ have to distributed between the physical qubits and the links between them (see Figure~\ref{fig:chain_biases}). The default method of D-Wave distributes $h_i$ and $J_{ij}$ uniformly between the qubits $\{x_i^1,\ldots,x_i^l\}$ and the links between the physical qubits (chains) implementing $x_i$ and $x_j$, respectively. However, there is no evidence that such a method is optimal. In fact, Pudenz~\cite{Pudenz16} has compared the default method with two other distribution strategies and has shown that in some cases the alternative methods work better.

However, none of the previous strategies considers the possible effect of hardware biases, or looks at bias distribution strategies to mitigate such issues. Here we address this problem, using our fixed embedding approach, and tackle the bias distribution problem (i.e., how to distribute the biases on the physical qubits and couplers in a way as to optimize the annealing results) as an optimization problem. The optimization of linear weights and quadratic couplers is done separately, denoted as CW(L) and CW(Q), respectively. When optimizing linear weights, we evenly distribute the quadratic weights among the quadratic couplers, and analogously for the linear weights when optimizing quadratic couplers.

\subsection{Formulations of the NP-hard problems studied}
\label{sec:np_problems}
We consider three classical NP-hard problems in this work, the Maximum Clique, Maximum Cut, and Graph Partitioning problems. For a graph $G=(V,E)$ with vertex set $V$ and edge set $E$, a \textit{clique} in $G$ is any subgraph $C$ of $G$ that is \textit{complete}, i.e., there is an edge between each pair of vertices of $C$. The \textit{Maximum Clique problem} asks us to find a clique of maximum size. A formulation of the Maximum Clique problem in the form of eq.~\eqref{eq:hamiltonian} can be found in \cite{Pelofske2019}.

Similarly, a \textit{cut} of the graph is any partition of $V$ into two disjoint sets, that is $V = V_1 \cup V_2$ and $V_1 \cap V_2 = \emptyset$. The \textit{cut size} of any cut is the number of edges having one endpoint in $V_1$ and one endpoint in $V_2$, that is $|\{ e=(v,w): v \in V_1, w \in V_2 \}|$. The \textit{Maximum Cut problem} asks us to find a cut of maximum size. A formulation of the Maximum Cut problem as an Ising problem can be found in \cite{github-maxcut}.

Last, the \textit{Graph Partitioning problem} asks us to divide the set of vertices $V$ into two disjoint and balanced sets (partitions) $V_1$ and $V_2$, satisfying $V = V_1 \cup V_2$ and $V_1 \cap V_2 = \emptyset$, such that the size of $V_1$ and $V_2$ differs by at most one and the number of \textit{cut edges} $\{ e=(v,w): v \in V_1, w \in V_2 \}$ between the two partitions is minimized. An Ising formulation for Graph Partitioning is given in \cite{github-gp}.

\subsection{Differential evolution optimization}
\label{sec:optimization}
In this work, we aim to tune three parameters per qubit, that is, its spin indicator for spin reversal, its anneal offset, and its chain weights in case we are dealing with a chained qubit on the D-Wave Chimera hardware.

To carry out the optimization we employ the \textit{differential optimization} solver of the \textit{SciPy} library in Python \cite{SciPy}, available under the command \textit{scipy.optimize.} \textit{differential\_evolution()}, which implements the algorithm of Storn and Price \cite{StornPrice1997}.

We employ the differential optimizer with a population size of $80$ and $50$ generations. We do not use the option \textit{polishing}, i.e., no steepest decent with a quasi-Newton method is performed to fine-tune the solution. All remaining parameters are left at their default values. The initial population is random, but we made sure to include default parameters for SR, AO, and CW given by D-Wave Systems, Inc. Lastly, we use elitism in the optimization, i.e., we make sure that the best solution is always passed on to the next generation, as opposed to the possibility of being replaced by crossover and random selection operations.

\section{Experimental analysis}
\label{sec:experiments}
We will perform the optimization on a class of random test graphs, separately for the three problems introduced in Section~\ref{sec:np_problems}, and evaluate the performance on a series of (unseen) testing graphs. With this, we aim to find out if it is possible to enhance the performance of the D-Wave 2000Q on a whole class of problem instances, since it is easy to see that due to the large search space, optimizing parameters for each newly solved problem is infeasible. 

In all experiments, we fix the anneal duration at $1000$ microseconds. We generate $10$ training and $10$ testing graphs, each with $65$ vertices, the size of the largest complete graph embeddable onto the D-Wave hardware. We vary the density of the training and validation graphs in $\{0.25,0.50,0.75\}$. We use the majority vote unembedding algorithm. For Maximum Clique and Maximum Cut we use a chain strength of 1, and for Graph Partitioning a chain strength of $20 \cdot 32 \cdot 33 \cdot d$, where $d$ is the graph density. This is similar to the choice in \cite{github-gp}, where the chain strength for Graph Partitioning is set to a prefactor multiplied with an estimate of the value of the objective function.

For the optimization, for each fixed point in the parameter search space, we perform $1000$ anneals per graph, and record the average performance across all $10$ training graphs, measured in both the value of the QUBO/Ising objective function and the energy (before unembedding) returned by the D-Wave annealer. For testing, we perform $10000$ anneals per graph.

When reporting the experimental results, we denote by \textit{Default-RE} the default behavior of the D-Wave annealer with a random embedding and with all other parameters set to their default values. As the optimization is performed on the same embedding, it is reasonable to try to find one that will result in the best performance on average. Hence, we try $30$ random embeddings and choose one for each problem that yields the best objective function value during forward annealing with default parameters (since Maximum Clique and Maximum Cut are maximization problems, the higher the value the better, whereas for Graph Partitioning, which is a minimization problem, lower is better). We denote this as \textit{Default-OE} (for default D-Wave with optimized embedding). Moreover, we denote by SR(Q) and SR(C) the tuned spin reversal on the qubit or chain level, and similarly AO(Q) and AO(C) denote the tuned anneal offsets on the qubit or chain level. Moreover, CW(L) and CW(Q) refer to setting the chain weights of linear or quadratic couplers. Since the D-Wave 2000Q features \textit{auto\_scale} and \textit{extended\_j\_range} are mutually exclusive, and because we use auto scaling for all experiments, we optimized CW(Q) without the extended J range feature.

We assess the performance of all methods using the time-to-solution metric, defined as the time to reach an optimum solution at least once with probability 0.99. It is computed as $\text{TTS} = T_\text{QPU} \cdot \log(0.01)/\log(1-p)$, where $T_\text{QPU}$ is the solve time on D-Wave, and $p$ is the proportion of times the optimal solution was found. Two caveats are worth mentioning: For problem instances where the optimal solution can be found using a classical solver, we are able to compute the time-to-optimal-solution, which we simply denote by TTS (time-to-solution). In case the optimal solution cannot be found in reasonable time, we relax this metric to \textit{time-to-best-solution}, denoted by TBS, which uses the best solution found by any of the methods, instead of the provably best one. The TBS measure depends on the set of algorithms employed in the study, their parameters, and the D-Wave samples on which these algorithms are run. However, the setup of the simulations presented here is fixed, thus making the TBS measure well-defined. Lastly, the TTS measure we report does not include any classical computation time, nor the training portion of the optimization done for each problem.

\begin{figure}[t]
    \centering
    \includegraphics[width=\textwidth]{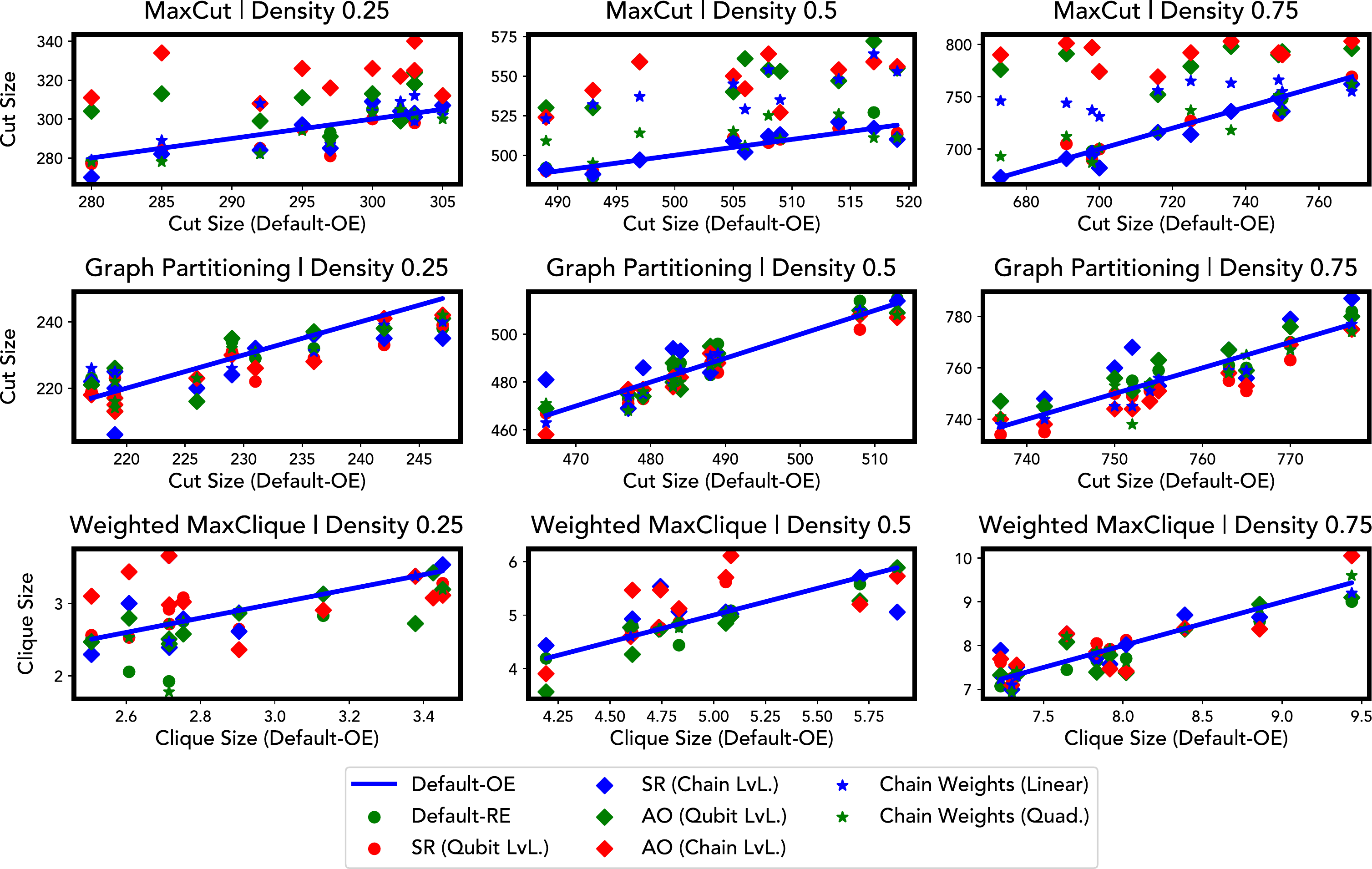}
    \caption{Performance on test graphs compared to Default-OE for graphs of density $0.25$ (left column), $0.5$ (middle column), and $0.75$ (right column). Metric is the cut size for the Maximum Cut problem (top row), cut size for the Graph Partitioning problem (middle row), and clique size for the Maximum Clique problem (bottom row).\label{fig:testgraph_performance}}
\end{figure}
Figure~\ref{fig:testgraph_performance} gives a graphical representation of our results on the suite of test graphs for all three problems. We measure results in the raw values of the found cut size (for Maximum Cut and Graph Partitioning) or clique size for the Maximum Clique problem. We observe that, compared to the baseline of Default-OE, SR as well as AO and CW seem to perform well for Maximum Cut. For the other two problems, it is mostly SR and AO on the chain level which perform best.

\begin{table}[t]
    \centering
    \footnotesize
    \begin{tabular}{|l|r|rr rr rr rr|}
        \hline
        Problem & Density & Default-OE & Default-RE & SR(Q) & SR(C) & AO(Q) & AO(C) & CW(L) & CW(Q)\\
        \hline
        \hline
        MaxCut  & 0.25 & & & & & 2857.8 & 5212.4 & \textbf{1121.0} &\\
                &&&&&& (1) & \textbf{(10)} & (1) &\\
                & 0.5 & & & & & 3246.2 & \textbf{3080.3} &&\\
                &&&&&& (5) & \textbf{(6)}&&\\
                & 0.75 & & & & & \textbf{1436.1} & 3185.2 &&\\
                &&&&&& (3) & \textbf{(9)} &&\\
        \hline
        GraphPart.  & 0.25 & 7179.0 & & 5809.4 & 3457.0 & 3869.6 & 4060.5 & & \textbf{3396.9}\\
                    && (1) && (3) & \textbf{(3)} & (1) & (2) && (1)\\
                    & 0.5 & & 7215.7 & 7987.9 & & 6817.3 & \textbf{3569.4} & 3733.9 & 3853.2\\
                    &&& (2) & \textbf{(4)} && (1) & (3) & (1) & (2)\\
                    & 0.75 & & & 7669.1 & & & 4286.7 & & \textbf{3961.7}\\
                    &&&& \textbf{(6)} &&& (3) && (2)\\
        \rowcolor{Gray}
        \hline
        MaxClique   & 0.25 & 14979.0 & & 11520.8 & \textbf{8226.6} & & & 8292.9 & 10728.6\\
                    \rowcolor{Gray}
                    && (3) && (1) & (2) &&& \textbf{(3)} & (2)\\
                    \rowcolor{Gray}
                    & 0.5 & 683.1 & 6061.0 & 2612.8 & 5571.9 & \textbf{111.7} & & 559.4 & 1754.3\\
                    \rowcolor{Gray}
                    && (2) & \textbf{(3)} & (2) & (2) & (1) && (2) & (2)\\
                    \rowcolor{Gray}
                    & 0.75 & 9.9 & 323.7 & \textbf{4.1} & 13825.2 & & & 58.6 & 20.2\\
                    \rowcolor{Gray}
                    && (1) & (1) & \textbf{(1)} & (1) &&& (1) & (1)\\
        \hline
    \end{tabular}
    \smallskip
    \caption{TBS (white) and TTS (gray) for the test graphs. Number of test graphs for which optimal solutions were found in parentheses. The best TTS and the highest number of optimal/best solutions found for each problem/density combination are given in bold. In case of a tie, the combination with the smallest TTS is chosen.\label{tab:TBSTTS}}
\end{table}
Next, the results for the TTS estimations are given in Table~\ref{tab:TBSTTS}. It is complicated to rank the performances of these techniques, since two measures are of relevance here. First, a low TTS or TBS time is desired for any method. However, some methods might be able to achieve a low TTS/TBS time, but only at the expense of solving fewer test graphs than others, and a method solving almost all graphs usually incurs a higher TTS/TBS time (this occurs, for instance, for the Maximum Cut problem and density 0.25). We thus denote both the measured TTS/TBS time for the graph that could be solved, as well as the number of the graph problems for which the best solution could be found, in parentheses.

First, we note that for the Maximum Cut and Graph Partitioning problems, classical solvers are unable to find the optimal solution, meaning we have to resort to the TBS measure. For Maximum Cut, neither of the Optimized, Random, SR(Q), SR(C) method could compute the best known solution for any graph. Solely the optimized anneal offset feature (AO on qubit and chain level), and the optimized chain weights (CW for linear weights), can solve some of the graphs and attain best TBS measures. Overall, AO(C), annealing offsets at the chain level, gives the best performance for Maximum Cut.

For the Graph Partitioning problem, the optimized spin reversal can solve most problems on average, but only at the expense of incurring large TBS times. Using optimized anneal offsets on the chain level, as well as optimized chain weights (for quadratic couplers) yields best TBS times, however again at the expense of only solving few graphs which can bias the results. Spin reversal at the qubit level (SR(Q)) seems to be performing the best for this type of problem.

For the Maximum Clique problem, we are able to solve problem instances classically to optimality using the function \textit{networkx.algorithms.clique.find\_cliques} in Python's Networkx package, and thus report TTS times. We observe that all methods can only solve around two of the $10$ test graphs, and that overall, SR(Q) and CW(Q) yield the best TTS times for the Maximum Clique problem.

\begin{table}[t]
    \centering
    \begin{tabular}{|l|r|r rr rr rr|}
        \hline
        Problem & Density & Default-RE & SR(Q) & SR(C) & AO(Q) & AO(C) & CW(L) & CW(Q)\\
        \hline
        \hline
         MaxCut & 0.25 & -0.3 & -1.0 & -0.7 & 4.2 & \textbf{8.8} & 0.8 & -1.0\\
                & 0.50 & 0.2 & -0.0 & 0.1 & \textbf{8.8} & 8.3 & 7.2 & 1.3\\
                & 0.75 & 0.0 & -0.3 & -0.7 & 9.0 & \textbf{9.9} & 4.4 & 0.1\\
        \hline
        GraphPart.\     & 0.25 & 0.3 & \textbf{1.6} & 1.3 & 0.6 & 1.4 & 0.2 & 0.6\\
                        & 0.50 & -0.2 & \textbf{0.6} & -0.6 & 0.0 & 0.4 & 0.1 & 0.5\\
                        & 0.75 & -0.2 & \textbf{0.7} & -0.7 & -0.4 & 0.6 & 0.4 & 0.2\\
        \hline
        MaxClique   & 0.25 & -8.8 & 0.5 & 2.3 & -4.6 & \textbf{6.4} & -1.8 & -6.6\\
                    & 0.50 & -0.6 & 1.2 & 2.3 & -3.3 & \textbf{5.5} & 0.3 & 0.3\\
                    & 0.75 & -2.2 & \textbf{3.1} & 1.8 & -1.4 & 0.3 & 0.1 & 0.2\\
        \hline
    \end{tabular}
    \bigskip
    \caption{Improvement (\%) in the value of the QUBO/Ising formulation compared to Default-OE for the test graphs.\label{tab:Improvement}\vspace{-0.5em}}
\end{table}
Apart from reporting TBS/TTS times, we can also evaluate all methods using the value of the Ising or QUBO formulation on the bitstring returned by D-Wave. Doing this for the best of the $30$ embeddings on D-Wave with default parameters (i.e., Default-OE) allows us to set a reference point, and we report the improvement (in percent) of the obtained value over this reference in Table~\ref{tab:Improvement}.

For the Maximum Cut problem, optimized anneal offsets (both on the qubit and the chain level) resulted in the largest percent improvement. For graph partitioning, spin reversal on the qubit level performs best, though the improvements are only of the order of one percent. Surprisingly, using Default-RE for Maximum Cut and Graph Partitioning does not perform very different than Default-OE. Lastly, for Maximum Clique, using anneal offsets on the chain level performs considerably better than the other techniques for density 0.25 and 0.5, with spin reversal being best for high densities.

\section{Discussion}
\label{sec:discussion}
This work considered optimizing three recent features of the D-Wave 2000Q annealer, precisely spin reversal (on qubit or chain level), anneal offsets (on qubit or chain level), and chain weight distribution (for linear or quadratic couplers). After fixing the embedding, we perform a classical optimization over a suite of random test graphs using a differential evolution optimizer, and aim to investigate if it is possible to outperform the default D-Wave anneal setting with optimized parameters for the three techniques. Our overall aim is to tune SR, AO and CW such that these features work better on a whole class of problems.

We conclude that for random graphs, and the three NP-hard problems we considered, tuning anneal offsets indeed works best, yielding substantial improvements over the default D-Wave behavior, especially for the Maximum Cut and Graph Partitioning problems. Optimizing spin reversal and chain weights seems more dependent on the problem and measure.

This work leaves scope for a variety of future research avenues. First, we performed the optimization for SR, AO and CW individually, since each involves tuning around $2000$ variables, and we found larger optimization problems to be infeasible. More elaborate optimization methods could allow us to optimize all parameters simultaneously, potentially improving results. Second, it would be interesting to extend the experiments to more classes of NP-hard problems, with the aim to see how much the trained parameters of SR, AO, or CW differ, and to investigate if the trained SR, AO, or CW can be recycled for certain classes. Third, the fixed embedding setup could allow us to determine if certain hardware qubits behave more favorably if consistently spin reversed, or if consistently employed with a particular anneal offset. Finally, though time consuming, our experiments would benefit from larger sets of testing and training graphs.

\section*{Acknowledgments}
This work has been supported by the US Department of Energy through the Los Alamos National Laboratory. Los Alamos National Laboratory is operated by Triad National Security, LLC, for the National Nuclear Security Administration of U.S. Department of Energy (Contract No.~89233218CNA000001) and by the Laboratory Directed Research and Development program of Los Alamos National Laboratory under project numbers 20190065DR and 20180267ER.


\end{document}